\documentclass[twocolumn]{autart}    
   
\usepackage[cmex10]{amsmath}
\usepackage{epsfig,amsfonts,amssymb,cite,times,subfigure,floatflt,wrapfig,enumerate,tikz,url,bm}
\usepackage{algorithmicx,algpseudocode}
\usepackage{booktabs}
\usepackage{siunitx}
\usepackage{graphicx}          

\newtheorem{lemma}{Lemma}

\newcommand{\mat}[1]{\bm{#1}}
\newcommand{\ten}[1]{\bm{\mathcal{#1}}}

\newcommand{\kpr}[1]{\textsuperscript{\textcircled{#1}}}

\begin{document}

\begin{frontmatter}

\title{Matrix output extension of the tensor network Kalman filter with an application in MIMO Volterra system identification}


\author[HKU]{Kim Batselier}\ead{kim.batselier@eee.hku.hk},    
\author[HKU]{Ngai Wong}\ead{nwong@eee.hku.hk},               

\address[HKU]{The Department of Electrical and Electronic Engineering, The University of Hong Kong}  


\begin{abstract}                          
This article extends the tensor network Kalman filter to matrix outputs with an application in recursive identification of discrete-time nonlinear multiple-input-multiple-output (MIMO) Volterra systems. This extension completely supersedes previous work, where only $l$ scalar outputs were considered. The Kalman tensor equations are modified to accommodate for matrix outputs and their implementation using tensor networks is discussed. The MIMO Volterra system identification application requires the conversion of the output model matrix with a row-wise Kronecker product structure into its corresponding tensor network, for which we propose an efficient algorithm. Numerical experiments demonstrate both the efficacy of the proposed matrix conversion algorithm and the improved convergence of the Volterra kernel estimates when using matrix outputs.
\end{abstract}

\end{frontmatter}

\section{Introduction}
In \cite{TNKalman}, a tensor network Kalman filter was developed to solve the state estimation problem of the following linear discrete-time state space model
\begin{align}
\nonumber \mat{X}(t+1) &= \mat{A}(t)\, \mat{X}(t) + \mat{W}(t),\\ 
\mat{y}(t) &= \mat{c}(t) \, \mat{X}(t) + \mat{r}(t),
\label{eqn:linearstatespace}
\end{align}
where $\mat{X}(t) \in \mathbb{R}^{n^d \times l}$ is a matrix containing $l$ exponentially long state vectors, $\mat{y}(t) \in \mathbb{R}^{1 \times l}$ is a row vector of $l$ scalar measurements, $\mat{A}(t) \in \mathbb{R}^{n^d \times n^d }$ is the state transition matrix, the row vector $\mat{c}(t) \in \mathbb{R}^{1 \times n^d}$ converts the state vectors into measurements and $\mat{W}(t) \in \mathbb{R}^{n^d \times l}, \mat{r}(t) \in \mathbb{R}^{1 \times l}$ denote zero-mean independent Gaussian process and measurement noise, respectively. Observe that unlike the conventional Kalman filter with a single state vector, \eqref{eqn:linearstatespace} represents a more general setting by concatenating $l$ state vectors into the matrix $\mat{X}(t)$ and $l$ scalar outputs into the row vector $\mat{y}(t)$. These $l$ state space models are then ``coupled" by a common state transition matrix $\mat{A}(t)$ and output model vector $\mat{c}(t)$. The simultaneous estimation of the $l$ state vectors is performed with one Kalman filter, whereby the standard Kalman equations are rewritten as tensor equations and implemented using tensor networks~\cite{TNorus,ivanTT}. The tensor network Kalman filter has two main advantages. First, the exponentially long mean vectors and covariance matrices never need to be explicitly formed. The second advantage is that the exponential storage and computational cost is transformed into a linear one.

The $l$ scalar output state space models~\eqref{eqn:linearstatespace} are motivated by their application in recursive MIMO Volterra system identification. Indeed, a discrete-time time-varying $p$-input $l$-output Volterra system of degree $d$ and memory $M$ is described by the state space model
\begin{align}
\nonumber \mat{X}(t+1) &= \mat{A}(t)\;\mat{X}(t)+ \mat{W}(t),\\ 
\mat{y}(t) &= \mat{u}_t\kpr{d} \; \mat{X}(t) + \mat{r}(t),
\label{eq:defMIMOVolterra}
\end{align}
where the row vector
\begin{align*}
\mat{u}_t := \begin{pmatrix}1 & u_1(t) & u_2(t) & \cdots & u_p(t-M+1) \end{pmatrix} \in \mathbb{R}^{1\times (pM+1)} 
\end{align*}
contains all $p$ input values at times $t$ down to $t-M+1$ and $\mat{u}_t\kpr{d}$ is defined as the $d$-times repeated Kronecker product
\begin{align}
\label{eqn:utkrp}
\mat{u}_t\kpr{d} := \overbrace { {\mat{u}_t\otimes \mat{u}_t\otimes \cdots \otimes \mat{u}_t}}^{d}\; \in\; \mathbb{R}^{1 \times (pM+1)^d}.
\end{align}
Each column of the $(pM+1)^d \times l$ matrix $\mat{X}(t)$ contains all coefficients from the Volterra kernels of degree $0$ up to degree $d$ for each of the $l$ corresponding outputs. A tensor network Kalman filter is ideally suited for recursive MIMO Volterra system identification, given the exponential size of the Volterra coefficients in $\mat{X}(t)$. The repeated Kronecker product form of equation~\eqref{eqn:utkrp} in particular lends itself well to a tensor network implementation as $\mat{u}_t\kpr{d}$ has a rank-1 tensor network representation~\cite[Lemma 4, p.~23]{TNKalman}, resulting in a significant reduction of storage cost and computational complexity.

However, the limitation to a row vector output $\mat{y}(t)$ is quite restrictive for the applicability of the tensor network Kalman filter to more generic dynamical systems. 
This provides the main motivation to extend the tensor network Kalman filter framework to the following state space model
\begin{align}
\nonumber \mat{X}(t+1) &= \mat{A}(t)\, \mat{X}(t) + \mat{W}(t),\\ 
\mat{Y}(t) &= \mat{C}(t) \, \mat{X}(t) + \mat{R}(t),
\label{eqn:linearstatespace2}
\end{align}
where now $\mat{Y}(t) \in \mathbb{R}^{m \times l}$, $\mat{C}(t) \in \mathbb{R}^{m \times n^d}$ and \mbox{$\mat{R}(t) \in \mathbb{R}^{m \times l}$} are matrices. The main contribution of this brief paper is twofold:
\begin{enumerate}
\item A constructive algorithm is proposed to convert the MIMO Volterra output model matrix $\mat{C}(t)$ into its corresponding tensor network. By exploiting the specific structure of the output model matrix $\mat{C}(t)$, a much more computationally efficient conversion is obtained, which is a crucial component for real-time identification of MIMO Volterra systems.
\item The Kalman tensor equations that appear in~\cite[p.~20]{TNKalman} are rewritten to accommodate for matrix outputs. This involves a nontrivial modification of the computations involved and their implementation using tensor networks is discussed.
\end{enumerate}
Numerical experiments in Section~\ref{sec:experiment} demonstrate the efficacy of our proposed conversion algorithm and compare the performance of the tensor network Kalman filter described in~\cite{TNKalman} with the newly proposed matrix output tensor network Kalman filter. It will be shown that using a matrix output can double the convergence speed of the estimated Volterra kernel coefficients at practically no additional cost. The matrix output Tensor Network Kalman filter proposed here therefore supersedes the work in~\cite{TNKalman}.

\section{Tensor notation}
Tensors in this article are multi-dimensional arrays that generalize the notions of vectors and matrices to higher orders. A $d$-way or $d$th-order tensor is denoted $\ten{A} \in \mathbb{R}^{n_1 \times  n_2 \times \cdots \times n_d}$ and hence each of its entries $\ten{A}(i_1,i_2,\cdots, i_d)$ is determined by $d$ indices. The numbers $n_1,n_2,\ldots,n_d$ are called the dimensions of the tensor. For practical purposes, only real tensors are considered. We use boldface capital calligraphic letters $\ten{A},\ten{B},\ldots$ to denote tensors, boldface capital letters~$\mat{A},\mat{B},\ldots$ to denote matrices, boldface letters~$\mat{a},\mat{b},\ldots$ to denote vectors, and Roman letters $a,b,\ldots$ to denote scalars. The transpose of a matrix $\mat{A}$ or vector $\mat{a}$ are denoted by $\mat{A}^T$ and $\mat{a}^T$, respectively. MATLAB colon notation is used to specify ``slices'' of tensors. For example, $\mat{A}(:,i)$ denotes the $i$th column of the matrix $\mat{A}$, while $\ten{A}(:,:,i)$ denotes the $i$th matrix slice of a third-order tensor $\ten{A}$. A more detailed description of the tensor concepts and operations used in this article can be found in~\cite{MVMALS,TNKalman}. 

\section{Converting the output model matrix $\mat{C}(t)$ into a tensor network}
The extension of the tensor network Kalman filter to matrix outputs requires the conversion of the output model matrix $\mat{C}(t)$ into a tensor network. Two cases will be considered. First, we briefly discuss the case where $\mat{C}(t)$ is a generic matrix. The second case deals specifically with the identification of MIMO Voltera systems, for which the $\mat{C}(t)$ matrix turns out to be highly structured. 

\subsection{Generic matrix $\mat{C}(t)$}
The two most common algorithms for the conversion of a matrix to its corresponding tensor network are the TT-SVD~\cite[p.~2301]{ivanTT} and the TT-cross approximation~\cite[p.~82]{ttcross} algorithms. The first step of applying the TT-SVD algorithm to the $\mat{C}(t)$ matrix consists of reshaping the matrix into an $mn \times n^{d-1}$ matrix and computing its singular value decomposition (SVD). This implies that the whole matrix $\mat{C}(t)$ needs to be kept in memory, which quickly becomes infeasible for increasing values of $n$ and $d$. The remaining steps of the TT-SVD algorithm are consecutive reshapings and SVDs of the obtained right singular vectors. The dominating computational step has a cost of approximately $O(14n^{d+1}m^2 +8m^3n^3)$ flops~\cite[p.~254]{matrixcomputations}. In addition to the storage cost, the exponential term $n^{d+1}$ in the computational complexity limits the applicability of the TT-SVD algorithm even further.

The TT-cross approximation algorithm circumvents these limitations by replacing the expensive SVD computation with another dyadic decomposition, the skeleton or pseudo-skeleton decomposition~\cite{GOREINOV19971}. The complexity of the TT-cross approximation algorithm has a linear dependence on $d$ and also works when the tensor entries are described by a function, thus eliminating the need to store the whole matrix in memory. While the TT-SVD algorithm is able to compute a tensor network representation of a given matrix with a machine precision accuracy, this is more difficult for the TT-cross algorithm. In addition, the TT-cross algorithm is usually slow as it is likely that it needs to be restarted when the desired accuracy is not met. A recent alternative for the conversion of a sparse matrix to a tensor network is described in~\cite{TNrSVD}. This newly-proposed algorithm converts a given sparse matrix directly into a tensor network without any dyadic decomposition. Instead, the sparse matrix is partitioned into block matrices, which can be explicitly written in tensor network form. Machine precision accuracy is guaranteed and runtimes are reported to be 500 times faster than the TT-SVD and TT-cross algorithms.

When $\mat{C}(t)$ is time-independent, the conversion of this matrix to its tensor network only needs to be done once. The storage and computational complexity of the conversion algorithm is then not that critical as the computation can be done prior to starting the Kalman filter. When $\mat{C}(t)$ is time-varying, then the conversion to its tensor network is required for each iteration of the Kalman filter. This imposes a severe restriction on the applicability of the tensor network Kalman filter for real-time filtering of generic time-varying dynamical systems. Note that the same argument also applies to the $\mat{A}(t)$ matrix. As we will demonstrate in the next subsection, the situation improves significantly when $\mat{C}(t)$ is structured, whereby it is possible to exploit the structure to efficiently obtain its tensor network representation.

\subsection{MIMO Volterra output model matrix $\mat{C}(t)$}
Before discussing the conversion of the MIMO Volterra output model matrix $\mat{C}(t)$ to a tensor network, we first need to extend the linear state space model for discrete-time MIMO Volterra systems~\eqref{eq:defMIMOVolterra} to the matrix output case. This extension is achieved by considering multiple output samples at once. These output samples can come from different experiments (e.g., with different applied inputs), or by grouping consecutive output measurements together. Without loss of generality, we consider the case where $m$ consecutive output measurements are concatenated to obtain the extended state space model
\begin{align}
\nonumber \mat{X}(t+1) &= \mat{A}(t)\;\mat{X}(t)+ \mat{W}(t),\\ 
\begin{pmatrix} \mat{y}(t) \\ \mat{y}(t+1) \\ \vdots \\ \mat{y}(t+m-1) \end{pmatrix} &= \begin{pmatrix} \mat{u}_t\kpr{d} \\ \mat{u}_{t+1}\kpr{d} \\ \vdots \\ \mat{u}_{t+m-1}\kpr{d} \end{pmatrix} \; \mat{X}(t) + \mat{R}(t).
\label{eq:defMIMOVolterra2}
\end{align}
Each row of the $\mat{C}(t)$ matrix consists of a $d$-times repeated Kronecker product, resulting in a highly structured dense matrix. Applying the TT-SVD to this matrix is not feasible for large values of $pM+1$ and $d$, due to the exponential storage and computational complexity. The TT-cross approximation algorithm on the other hand is too slow in order to deploy it for real-time applications. Finally, the alternative matrix to tensor network conversion algorithm reported in~\cite{TNrSVD} is best suited for sparse matrices, while $\mat{C}(t)$ will contain many nonzero entries. Fortunately, it is possible to derive an efficient algorithm that exploits the repeated Kronecker product structure to construct an exact tensor network representation of $\mat{C}(t)$. Before providing the constructive derivation of the main algorithm, we first revisit the result for the row vector $\mat{c}(t)$ as described in~\cite{TNKalman} and explain why it fails for the matrix output case.
\subsection{Failure of row vector result to the matrix case}
The repeated Kronecker product structure of \mbox{$\mat{c}(t)=\mat{u}_t\kpr{d}$ in~\eqref{eq:defMIMOVolterra}} gives rise to the following tensor network.
\begin{lemma}(Lemma 4 of~\cite[p.~23]{TNKalman})
The unit-rank tensor network of $\mat{u}_t\kpr{d}$ consists of $d$ tensors $\ten{U}^{(k)} \in \mathbb{R}^{1 \times (pM+1) \times 1}$ with $k=1,\ldots,d$ and $\ten{U}^{(k)}(1,:,1) := \mat{u}_t.$
\label{lemma:ut}
\end{lemma}
The tensor network of $\mat{c}(t)$ therefore consists of the vector $\mat{u}_t$ repeated $d$ times, with a total storage cost of $O(pM+1)$ since we only need to store the vector $\mat{u}_t$ once. Now, defining the $m \times (pM+1)$ matrix $\mat{U}_t$ as
\begin{align*}
\mat{U}_t &:= \begin{pmatrix} \mat{u}_t\\
\mat{u}_{t+1}\\ \vdots \\ \mat{u}_{t+m-1} \end{pmatrix},
\end{align*}
we can rewrite the matrix $\mat{C}(t)$ in~\eqref{eq:defMIMOVolterra2} as
\begin{align}
\label{eqn:Utkrp}
 \begin{pmatrix} \mat{u}_t\kpr{d} \\ \mat{u}_{t+1}\kpr{d} \\ \vdots \\ \mat{u}_{t+m-1}\kpr{d} \end{pmatrix} &= \overbrace { {\mat{U}_t \odot \mat{U}_t \odot  \cdots \odot \mat{U}_t}}^{d},
\end{align}
where $\odot$ denotes the row-wise Kronecker product. Note that in~\cite{TNKalman} the notation $\odot$ is used to denote the column-wise Kronecker product.  If $\mat{C}(t)$ in~\eqref{eq:defMIMOVolterra2} had been the $d$-times repeated Kronecker product of $\mat{U}_t$, then Lemma \ref{lemma:ut} would also apply, resulting in a simple rank-1 tensor network where each of the $d$ tensors is $\ten{U}^{(k)} \in \mathbb{R}^{1 \times m\times (pM+1) \times 1}$ and $\ten{U}^{(k)}(1,:,:,1):=\mat{U}_t$. Finding the tensor network representation $\ten{U}^{(1)},\ldots,\ten{U}^{(d)}$ of a matrix with a row-wise Kronecker product structure, however, is a nontrivial problem. A constructive algorithm that solves this problem is proposed in the next section.

\section{Constructive algorithm}
The main idea of our proposed algorithm is to start with the computation of the last tensor $\ten{U}^{(d)}$ and build up the whole network one tensor at a time. The first step is to compute the row-wise Kroncker product $\mat{U}_t \odot \mat{U}_t$, which results in an $m \times (pM+1)^2$ matrix. This matrix is then reshaped into an $m(pM+1) \times (pM+1)$ matrix $\mat{T}$ and its SVD is
\begin{align}
\mat{T} &= \mat{U}\; \mat{S} \; \mat{V}^T,
\label{eqn:svdT}
\end{align}
where $\mat{U}\in \mathbb{R}^{m(pM+1)\times r_{d-1}}, \mat{V} \in \mathbb{R}^{(pM+1)\times  r_{d-1}}$ are orthogonal matrices, $\mat{S} \in \mathbb{R}^{ r_{d-1} \times  r_{d-1}}$ is a diagonal matrix with nonnegative entries and $r_{d-1}$ is the numerical rank of $\mat{T}$. The tensor $\ten{U}^{(d)}$ is completely determined by reshaping the matrix product $\mat{S}\mat{V}^T$ into a $ r_{d-1} \times (pM+1) \times 1$ tensor. In order to compute $\ten{U}^{(d-1)}$, one can repeat the same procedure with the left singular vectors $\mat{U}$ of~\eqref{eqn:svdT}. The matrix $\mat{U}$ is reshaped into an $m\times (pM+1) r_{d-1}$ matrix and another row-wise Kronecker product with $\mat{U}_t$ is computed. The resulting matrix is also factored with an SVD and the tensor $\ten{U}^{(d-1)}$ is retrieved from the reshaped matrix product $\mat{S}\mat{V}^T$. The left singular vectors are then again reshaped and the procedure repeats until the first tensor $\ten{U}^{(1)}$ of the network is found. The whole algorithm is presented in pseudo-code as Algorithm~\ref{alg:matrix2TN}.

The correctness of the algorithm is easily verified as it consists of computing the desired row-wise Kronecker products in~\eqref{eqn:Utkrp} with intermediate SVD computations. The most computationally expensive step in Algorithm~\ref{alg:matrix2TN} is the SVD of the $\mat{T}$ matrix. Assuming for notational convenience that $r_1=r_2=\cdots = r_{d-1}=r$ and defining $n:=pM+1$, then the SVD of the $mn\times nr$ matrix $\mat{T}$ requires $O(14mn^3r^2+8r^3n^3)$ flops and is computed $d-1$ times. Compared to the TT-SVD algorithm, Algorithm~\ref{alg:matrix2TN} does not suffer from any exponential computational complexity. In addition, the obtained tensor network is guaranteed to be accurate up to machine precision when the SVD factorizations of $\mat{T}$ are not truncated. Note that Algorithm~\ref{alg:matrix2TN} is easily generalized to the case where each of the factors in the row-wise Kronecker product~\eqref{eqn:Utkrp} is a different matrix.
\begin{alg}Row-wise Kronecker product matrix to tensor network conversion\\
\label{alg:matrix2TN}
\textit{\textbf{Input}}: matrix $\mat{U}_t$, factor $d$\\
\textit{\textbf{Output}}:\makebox[0pt][l]{ tensor network $\ten{U}^{(1)},\ldots,\ten{U}^{(d)}$ of \eqref{eqn:utkrp}}
\begin{algorithmic}
\State $\ten{U}^{(d)} \gets \textrm{reshape}(\mat{U}_t,[1,m,pM+1])$
\For{$i=d,d-1,\ldots,2$}
\State $\mat{T} \gets \textrm{reshape}(\ten{U}^{(i)},[m,(pM+1)r_{i}])$ \hfill \%\,$r_d=1$
\State $\mat{T} \gets \mat{U}_t \odot \ \mat{T}$
\State $\mat{T} \gets \textrm{reshape}(\mat{T},[m(pM+1),(pM+1)r_i])$
\State $[\mat{U},\mat{S},\mat{V}] \gets \textrm{SVD}(\mat{T})$
\State $r_{i-1} \gets $ numerical rank of $\mat{T}$ determined from SVD
\State $\ten{U}^{(i)} \gets \textrm{reshape}(\mat{S}\mat{V}^T,[r_{i-1},1,pM+1,r_i])$
\State $\ten{U}^{(i-1)} \gets \textrm{reshape}(\mat{U},[1,m,pM+1,r_{i-1}])$
\EndFor
\end{algorithmic}
\end{alg}

\section{Modified Kalman tensor equations}
\label{sec:kalman}
The Kalman tensor equations described in~\cite[p.~20]{TNKalman} are only valid for $l$ scalar output state space models and therefore need to be modified in order to work for the matrix output case. First, we briefly review and extend the assumptions of the original tensor network Kalman filter, where we will continue to use the shorthand notation $n:=pM+1$:
\begin{itemize}
\item Each column $\mat{x}_k \,(k=1,\ldots,l)$ of the matrix $\mat{X}(t)$ follows a multivariate Gaussian distribution
\begin{align*}
\frac{1}{Z}\, \textrm{exp}\left( -\frac{1}{2} (\mat{x}_k-\mat{m}_k)^T \, \mat{P}_k \, (\mat{x}_k-\mat{m}_k) \right ),
\end{align*}
with normalization constant $Z:=((2\,\pi)^{n^d/2}\,|\mat{P}_k|^{1/2})$, where $|\mat{P}_k|$ denotes the determinant of $\mat{P}_k$. The vectors $\mat{m}_k$ are collected in the matrix $\mat{M}(t) \in \mathbb{R}^{n^d \times l}$ and similarly all covariance matrices are collected into a 3-way tensor $\ten{P}(t) \in \mathbb{R}^{n^d \times n^d \times l}$,
\item each column of the process noise matrix $\mat{W}(t)$ is a multivariate Gaussian white noise process. This implies zero means and diagonal covariance matrices, which are collected into a 3-way tensor $\ten{W} \in \mathbb{R}^{n^d \times n^d \times l}$,
\item likewise, each column of the measurement noise matrix $\mat{R}(t)$ is a multivariate Gaussian white noise process with zero means and diagonal covariance matrices, which are collected into a 3-way tensor $\ten{R} \in \mathbb{R}^{m \times m \times l}$,
\item The process noise $\mat{W}(t)$ and measurement noise $\mat{R}(t)$ are uncorrelated.
\end{itemize}
The additional assumption $m,l \ll n^d$ is made for practical considerations. Just as in~\cite[p.~20]{TNKalman}, the initial matrix $\mat{M}(0)$ is initialized to a rank-1 tensor network of all zeros. Each of the $l$ covariance matrices inside $\ten{P}(0)$ is a diagonal matrix with a constant value on the diagonal. This assumption also reduces the corresponding tensor network to be rank-1. We now go over each of the Kalman tensor equations and discuss the required modifications and tensor network implementations.
\subsection{Prediction steps}
Both prediction steps
\begin{align*}
\mat{M}^+ &= \mat{M}(t) \times_1 \mat{A}(t),\\
\ten{P}^+ &= \ten{P}(t) \times_1 \mat{A}(t) \times_2 \mat{A}(t) + \ten{W},
\end{align*}
remain unchanged. Their implementation using tensor networks is therefore as in~\cite[p.~20-21]{TNKalman}.
\subsection{$\mat{v} = \mat{y}(t) - \mat{M}^+ \times_1 \mat{c}(t)$}
With the extension to matrix outputs, the first update step needs to be modified into
\begin{align*}
\mat{V} = \mat{Y}(t) - \mat{M}^+ \times_1 \mat{C}(t),
\end{align*}
with $\mat{V} \in \mathbb{R}^{m \times l}$. The product $\mat{M}^+ \times_1 \mat{C}(t)$ is computed by contracting their respective tensor networks with each other. The tensor network for $\mat{C}(t)$ is obtained from Algorithm~\ref{alg:matrix2TN}. The assumption $m,l \ll n^d$ implies that the resulting matrix $\mat{V}$ is small enough to be stored in memory.
\subsection{$\mat{s} = \ten{P}^+\times_1 \mat{c}(t) \times_2 \mat{c}(t) + \textrm{diag}(\mat{R}(t))$}
The second update step changes quite significantly. The $l$-dimensional vector $s$ is now replace by the $m\times m \times l$ tensor $\ten{S}$, which is obtained from
\begin{align*}
\ten{S} &= \ten{P}^+\times_1 \mat{C}(t) \times_2 \mat{C}(t) + \ten{R}.
\end{align*}
The contraction $\ten{P}^+\times_1 \mat{C}(t) \times_2 \mat{C}(t)$ with tensor networks is performed in an identical way as the contraction $\ten{P}(t) \times_1 \mat{A}(t) \times_2 \mat{A}(t)$ from the prediction step. The resulting tensor network can then be contracted into an $m\times m \times l$ tensor and added to $\ten{R}$ directly.

\subsection{$\mat{K} = \ten{P}^+\times_2 \mat{c}(t) \times_3 \textrm{diag}(\mat{s})^{-1}$}
Computation of the Kalman gain also changes significantly. Whereas in~\cite{TNKalman} the Kalman gain is an $n^d \times l$ matrix, it now becomes an $n^d \times m \times l$ tensor $\ten{K}$ and is also stored as a tensor network. The contraction $\ten{P}^+\times_2 \mat{C}(t)$ is quite straightforward using tensor networks and results in an $n^d \times m \times l$ tensor. In the output matrix case the vector $\mat{s}$ is replaced by a tensor $\ten{S}$, which means that the scaling operation $\times_3 \textrm{diag}(\mat{s})^{-1}$ also needs to change. Each of the $l$ matrix slices $\ten{S}(:,:,i)$ is inverted and then contracted with each of the $l$ tensor slices of the first Kalman gain tensor network core
\begin{align}
\ten{K}^{(1)}(i,:,:,:) \times_2 \ten{S}(:,:,i)^{-1} \; \textrm{for} \; i=1,\ldots,l.
\label{eqn:kalmanscaling}
\end{align}
Observe that by fixing the first index in $\ten{K}^{(1)}(i,:,:,:)$ we obtain an $n \times m \times r_1$ tensor. Likewise, by fixing the last index in $\ten{S}(:,:,i)$ we obtain an $m \times m$ matrix.

\subsection{$\mat{M}(t+1) = \mat{M}^+ + \mat{K}\times_2 \textrm{diag}(\mat{v})$}
With the $\mat{v}$ vector replaced by the $m\times l$ matrix $\mat{V}$, the scaling operation $\mat{K}\times_2 \textrm{diag}(\mat{v})$ is computed in a similar fashion as in~\eqref{eqn:kalmanscaling}. Each of the $l$ tensor slices $\ten{K}^{(1)}(i,:,:,:)$ is contracted with each column of $\mat{V}$ as $\ten{K}^{(1)}(i,:,:,:)\times_2 \mat{V}(:,i)^T$ for $i=1,\ldots,l$. The resulting tensor network then corresponds with an $n^d \times l$ tensor, which is added to $\mat{M}^+$ in tensor network form to obtain the new estimate of the vector means $\mat{M}(t+1)$.

\subsection{$\ten{P}(t+1) = \ten{P}^+ - (\mat{K} \, \square \, \mat{K})\times_3 \textrm{diag}(\mat{s})$}
In the matrix output case, the Kalman gain becomes an $n^d\times m \times l$ tensor $\ten{K}$ for which the column-wise outer product operation $\square$, c.f.~\cite[p.~18]{TNKalman}, is not defined. Once again, the outer product will be computed for each of the $l$ slices $\ten{K}^{(1)}(i,:,:,:)$. We first define the $r_1n\times m$ matrix $\mat{K}_i$ as the matrix obtained from permuting and reshaping the $n \times m \times r_1$ tensor $\ten{K}^{(1)}(i,:,:,:)$. The matrix product
\begin{align*}
\mat{K}_i \, \ten{S}(:,:,i) \, \mat{K}_i^T
\end{align*}
then results in an $r_1n \times nr_1$ matrix, which is then permuted and reshaped into an $n \times n \times r_1^2$ tensor $\ten{K}_i$. This tensor $\ten{K}_i$ is then the $i$th slice of the tensor network core of $(\mat{K} \, \square \, \mat{K})\times_3 \textrm{diag}(\mat{s})$. Just as in Lemma 3 from~\cite[p.~22]{TNKalman}, the remaining tensor network cores of $(\mat{K} \, \square \, \mat{K})\times_3 \textrm{diag}(\mat{s})$ are the tensor Kronecker products $\ten{K}^{(k)} \otimes \ten{K}^{(k)} \, (k=2,\ldots,d)$. The resulting tensor network is then subtracted from the tensor network that represents $\ten{P}^+$ to obtain the updated covariance tensor $\ten{P}(t+1)$.

\section{Experiments}
\label{sec:experiment}
In this section, we demonstrate the effectiveness of Algorithm \ref{alg:matrix2TN} and the modified matrix output tensor network Kalman filter through numerical experiments. All computations were performed in MATLAB on an Intel i5 8-core processor running at 3.4 GHz with 64 GB RAM\footnote{MATLAB implementations of both Algorithm \ref{alg:matrix2TN} and the matrix output tensor network Kalman filter are freely available from \url{https://github.com/kbatseli/TNKalman}.}. 

\subsection{Converting $\mat{C}(t)$ into a tensor network}
In order to demonstrate the efficacy of Algorithm~\ref{alg:matrix2TN}, a matrix $\mat{U}_t \in \mathbb{R}^{100 \times 10}$ was created with samples drawn from a standard normal distribution. A conversion of the $\mat{C}(t)$ matrix with a repeated row-wise Kronecker product structure to its corresponding tensor network was performed with both the TT-SVD algorithm and Algorithm~\ref{alg:matrix2TN}. The TT-SVD algorithm was also implemented by ourselves in MATLAB. The conversion was run for $d=2,\ldots,7$ and 20 runs were performed for each value of $d$. The median runtime for each value of $d$ for both algorithms is shown in Table~\ref{tab:ex1}. Note that for $d=2$, the TT-SVD algorithm and Algorithm~\ref{alg:matrix2TN} are identical, as they both consist of computing the row-wise Kronecker product and a singular value decomposition. This explains the similar runtime for low values of $d$. The lower computational complexity of Algorithm~\ref{alg:matrix2TN} becomes more prominent for increasing values of $d$. For $d=7$, Algorithm~\ref{alg:matrix2TN} was able to compute the desired result at machine precision accuracy about 350 times faster than the TT-SVD algorithm.
\begin{table}[tb]
\begin{center}
\caption{Median runtime to convert $\mat{C}(t)$ into a tensor network with the TT-SVD and Algorithm \ref{alg:matrix2TN} for different values of $d$.}
\label{tab:ex1}	
\begin{tabular}{@{}lcc@{}}
$d$ & TT-SVD[s] & Algorithm~\ref{alg:matrix2TN}[s] \\ \midrule
2 & $0.0003$ &$0.0005$\\
3 & $0.0030$ &$0.0029$\\
4 & $0.1448$ &$0.0533$ \\
5 & $1.2025$ &$0.1720$\\
6 & $15.1880$&$0.3135$\\
7 & $157.676$ & $0.4414$\\
 \midrule
\end{tabular}
\end{center}
\end{table}

\subsection{Comparison of scalar with matrix output}
\begin{figure}[tb]
\begin{center}
\includegraphics[width=.5\textwidth]{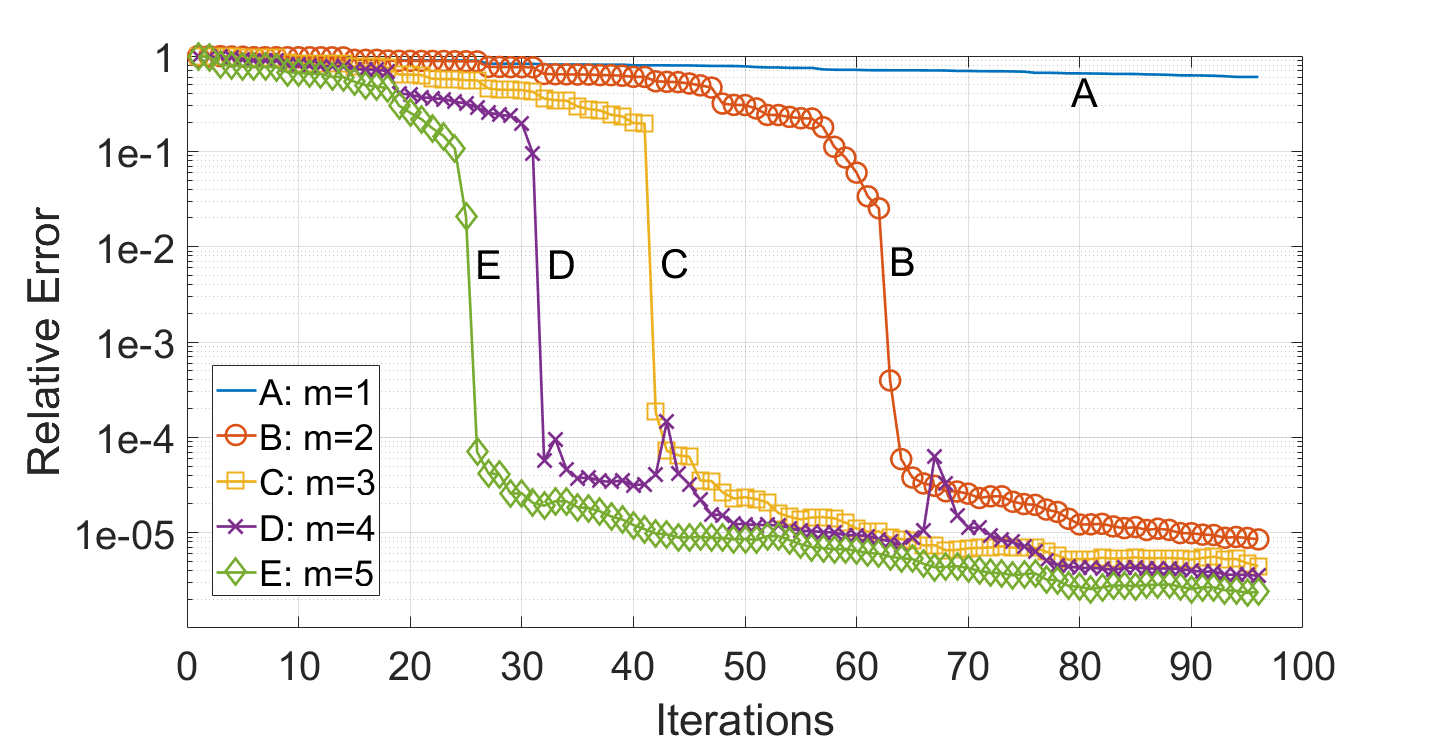}
\caption{Relative error of estimated Volterra coefficients for increasing number of output samples $m$.}
\label{fig:ex2}
\end{center}
\end{figure}

In order to be able to compare the performance of the scalar output Kalman filter with the matrix output Kalman filters in about 100 iterations, we consider the following time-invariant SISO Volterra system
\begin{align*}
\mat{x}(t+1) &= \mat{x}(t),\\
\mat{y}(t) &= \mat{C}(t)\, \mat{x} + \mat{r}(t),
\end{align*}
with $d=4, M=5$. The state vector $\mat{x}(t) \in \mathbb{R}^{1296}$ containing the Volterra kernel coefficients was constructed as $\mat{h}\kpr{d}$, where $\mat{h} \in \mathbb{R}^{6}$ is a random vector, sampled from a standard normal distribution. The measurement noise $\mat{r}(t)$ was sampled from a zero-mean Gaussian distribution with a variance of $10^{-8}$. All input samples were drawn from a standard normal distribution. A 100 iterations of tensor network Kalman filters were run for incremental values of $m$, starting from $m=1$ (the scalar output case) up to $m=5$. The covariance matrix was initialized to the identity matrix and a tolerance of $10^{-10}$ was set for the TN-rounding procedure. The relative error
\begin{align*}
\frac{||\mat{h}\kpr{d} - \mat{m}(t)||_2}{||\mat{h}\kpr{d}||_2}
\end{align*}
was computed at each iteration of the different tensor network Kalman filters and is shown in Figure \ref{fig:ex2}. Note that curve A in Figure~\ref{fig:ex2} corresponds with the result of the tensor network Kalman filter described in~\cite{TNKalman}. We can deduce from Figure \ref{fig:ex2} that a tensor network Kalman filter that processes $m$ output values per iteration converges approximately $m$ times faster. For example, the estimated state vector of the tensor network Kalman filter with $m=2$ reaches an accuracy of approximately 4 correct digits after about 65 iterations. The estimate of the tensor network Kalman filter with $m=4$ reaches the same accuracy after about 32 iterations. This comes as no surprise, as the amount of ``information" that is used to update the mean vector and covariance matrix increases proportional with the output size $m$. The median runtime per iteration and the total runtime for the different tensor network Kalman filters to obtain estimates with an accuracy of $10^{-4}$ are shown in Table~\ref{tab:ex2}. Remarkably, setting $m=2$ effectively doubles the convergence rate of the estimated Volterra coefficients at practically zero additional cost. Higher values of $m$ result in longer runtimes per iteration and longer total runtimes to reach a certain accuracy. This is entirely due to the tensor network rounding procedure, which is required to keep the network ranks small. In fact, 90\% of the runtime per iteration is spent in rounding, while the remaining 10\% are the actual Kalman tensor computations described in Section \ref{sec:kalman}. Future improvements in the computational complexity of the rounding procedure will therefore have an immediate benefit on the proposed tensor network Kalman filter.
\begin{table}[tb]
\begin{center}
\caption{Median runtime per iteration and total runtime to reach an accuracy of $10^{-4}$ for different values of $m$.}
\label{tab:ex2}	
\begin{tabular}{@{}lccccc@{}}
$m$		 &  1 & 2 & 3 & 4 & 5 \\ \midrule
$\mathrm{t}_\mathrm{median}$ per iteration [s]& 0.44 & 0.58 & 1.95 & 3.15 & 4.71  \\
$\mathrm{t}_\mathrm{total,1e-4}$ [s]& NA & 32.82 & 41.87 & 51.89 & 63.21  \\
 \midrule
\end{tabular}
\end{center}
\end{table}

\section{Conclusions}
This article presented an extension of the tensor network Kalman filter to matrix outputs with an application in the recursive identification of discrete-time nonlinear MIMO Volterra systems. The extension to matrix outputs completely supersedes the work reported in ~\cite{TNKalman}. A constructive algorithm was proposed that is able to efficiently convert the output model matrix $\mat{C}(t)$ with a row-wise Kronecker product structure into its corresponding tensor network. In addition, the Kalman tensor equations were modified to the matrix output case and their implementation using tensor networks were discussed. A possible future improvement is to enhance numerical stability of the computations through the implementation of a square-root tensor network Kalman filter.

\bibliographystyle{plain}        
\bibliography{references}           

\end{document}